\documentstyle[multicol,aps,epsf,amsmath]{revtex}
\begin{document}
\draft
\def\av#1{\langle#1\rangle}
\def\pc{p_{\rm c}}
\title {Resilience of the Internet to random breakdowns}
\author{Reuven~Cohen$^{1}$\footnote{{\bf e-mail:} cohenr@shoshi.ph.biu.ac.il},
 Keren~Erez$^1$, Daniel~ben-Avraham$^2$, and Shlomo~Havlin$^1$} 
\address{$^1$Minerva Center and Department of Physics, 
Bar-Ilan University, Ramat-Gan 52900, Israel}
\address{$^2$Physics Department and Center for Statistical Physics (CISP), 
Clarkson University, Potsdam NY 13699-5820, USA}
\maketitle
\begin{abstract}
A common property of many large networks, including the Internet, is that the 
connectivity of the various nodes follows a scale-free power-law distribution, 
$P(k)=ck^{-\alpha}$. We study the stability of such networks with respect to 
crashes, such as random removal of sites. Our approach, based on percolation 
theory, leads to a general condition for the critical fraction of nodes, 
$\pc$, that need to be removed before the network disintegrates. We show 
analytically and numerically that for $\alpha\leq 3$ the 
transition never takes place, unless the network is finite. In the special 
case of the physical structure of the Internet $(\alpha\approx 2.5)$, we find 
that it is  impressively robust, with $\pc > 0.99$.
\end{abstract}
\pacs{02.50.Cw, 05.40.a, 05.50.+q, 64.60.Ak}

\begin{multicols}{2}
\noindent
Recently there has been increasing interest in the formation of random networks
 and in the connectivity of these networks, especially in the context of the 
Internet \cite{albert2,Paxon,Zegura,Bar,albert,Huber,Claffy,redner,newman}. 
When such networks are subject to random breakdowns --- a fraction $p$ of the 
nodes and their connections are removed randomly --- their integrity might be 
compromised: when $p$ exceeds a certain threshold, $p>\pc$, the network 
disintegrates into smaller, disconnected parts. Below that critical threshold,
 there still exists a connected cluster that spans the entire system (its size
 is proportional to that of the entire system). Random breakdown in networks 
can be seen as a case of infinite-dimensional percolation. Two cases that have
 been solved exactly are Cayley trees \cite{essam} and Erd\H{o}s-R\'enyi (ER) 
random graphs \cite{bal}, where the networks collapse at known thresholds 
$\pc$. Percolation on small-world networks (i.e., networks where every node is
 connected to its neighbors, plus some random long-range connections~
\cite{watts}) has also been studied by Moore and Newman \cite{moore}. Albert 
{\it et al.}, have raised the question of random failures and intentional 
attack on networks~\cite{albert2}. Here we consider random breakdown in the 
Internet (and similar networks) and introduce an analytical approach to finding
 the critical point. The site connectivity of the physical structure of the 
Internet, where each communication node is considered as a site, is power-law,
to a good approximation~\cite{fal}. We introduce a new general criterion for 
the percolation critical threshold of randomly connected networks. Using this criterion, we show analytically that the Internet undergoes no transition under
 random breakdown of its nodes. In other words, a connected cluster of sites 
that spans the Internet survives even for arbitrarily large fractions of 
crashed sites. 

We consider networks whose nodes are connected randomly to each other, so that 
the probability for any two nodes to be connected depends solely on their 
respective connectivity (the number of connections emanating from a node). We 
argue that, for randomly connected networks with connectivity distribution 
$P(k)$, the critical breakdown threshold may be found by the following 
criterion: if loops of connected nodes may be neglected, the percolation 
transition takes place when a node ($i$), connected to a node ($j$) in the 
spanning cluster, is also connected to at least one other node --- otherwise 
the spanning cluster is fragmented. This may be written as
\begin{equation}
\label {criterion}
\av{k_i|i\leftrightarrow j}=\sum_{k_i}k_i P(k_i|i\leftrightarrow j)=2,
\end {equation}
where the angular brackets denote an ensemble average, $k_i$ is the 
connectivity of node $i$, and $P(k_i|i\leftrightarrow j)$ is the conditional 
probability  that node $i$ has connectivity $k_i$, given that it is connected 
to node $j$. But, by Bayes rule for conditional probabilities 
$P(k_i|i\leftrightarrow j)=P(k_i,i\leftrightarrow j)/P(i\leftrightarrow j)=
P(i\leftrightarrow j|k_i)P(k_i)/P(i\leftrightarrow j)$, where 
$P(k_i,i\leftrightarrow j)$ is the {\it joint\/} probability that node $i$ has 
connectivity $k_i$ and that it is connected to node $j$. For randomly 
connected networks (neglecting loops) $P(i\leftrightarrow j)=\av{k}/(N-1)$ and
 $P(i\leftrightarrow j|k_i)=k_i/(N-1)$, where $N$ is the total number of nodes 
in the network. It follows that the criterion~(\ref{criterion}) is equivalent 
to 
\begin{equation}
\label{kappa}
\kappa\equiv{\av{k^2}\over \av{k}}=2,
\end {equation}
at criticality.

Loops can be ignored below the percolation transition, $\kappa<2$, because 
the probability of a bond to form a loop in an $s$-nodes cluster is proportional to $(s/N)^2$ (i.e., proportional to the probability of choosing two sites in that cluster). The fraction of loops in the system $P_{loop}$ is
\begin{equation}
P_{loop}\propto \sum_i {s_i^2\over N^2} < \sum_i {s_i S\over N^2} = {S\over N},
\end {equation}
where the sum is taken over all clusters, and $s_i$ is the size of the $i$th 
cluster.  Thus, the overall fraction of loops in the system is smaller than 
$S/N$, where $S$ is the size of the largest existing cluster. Below criticality
 $S$ is smaller than order $N$ (for ER graphs $S$ is of order 
$\ln N$~\cite{bal}), so the fraction of loops becomes negligible in the limit 
of $N\rightarrow\infty$. Similar arguments apply at criticality. 

Consider now a random breakdown of a fraction $p$ of the nodes. This would 
generically alter the connectivity distribution of a node. Consider indeed a 
node with initial connectivity $k_0$, chosen from an initial distribution 
$P(k_0)$. After the random breakdown the distribution of the new connectivity 
of the node becomes $\binom{k_0}{k}(1-p)^k p^{k_0-k}$, and the new distribution
 is 
\begin{equation}
P'(k)=\sum_{k_0=k}^\infty P(k_0)\binom{k_0}{k}(1-p)^k p^{k_0-k}.
\end {equation}
(Quantities after the breakdown are denoted by a prime.) Using this new 
distribution one obtains $\av{k}'=\av{k_0}(1-p)$ and
$\av{k^2}'=\av{k_0^2}(1-p)^2+\av{k_0}p(1-p)$, so the criterion~(\ref{kappa}) 
for criticality may be re-expressed as
\begin{equation}
{\av{k_0^2}\over \av{k_0}}(1-\pc)+\pc=2,
\end {equation}
or
\begin{equation}
\label{perc}
1-\pc={1 \over \kappa_0-1},
\end {equation}
where $\kappa_0=\av{k_0^2}/\av{k_0}$ is computed from the original distribution
, before the random breakdown. 

Our discussion up to this point is general and  applicable to all randomly 
connected networks, regardless of the specific form of the connectivity 
distribution (and provided that loops may be neglected). 
For example, for random (ER) networks, which possess a Poisson connectivity
distribution, the criterion (\ref{kappa}) reduces to a known result\cite{bal} 
that the transition takes place at $\av{k}=1$. In this case, random breakdown 
does not alter the Poisson character of the distribution, but merely shifts its
 mean. Thus, the new system is again an ER network, but with new 
{\it effective} parameters: $k_{\rm eff} =k(1-p)$, $N_{\rm eff}=N(1-p)$. In the
 case of Cayley trees, the criteria (\ref{kappa}) and (\ref {perc}) also yield 
the known exact results~\cite{essam}. 

The case of the Internet is thought to be different. It is widely believed that
, to a good approximation, the connectivity distribution of the Internet nodes
 follows a power-law~\cite{fal}:
\begin{equation}  
\label {SF}
P(k)=ck^{-\alpha}, \quad k=m,m+1,...,K,
\end {equation}
where $\alpha\approx 5/2$, $c$ is an appropriate normalization constant, and 
$m$ is the smallest possible connectivity. In a finite network, the largest 
connectivity, $K$, can be estimated from
\begin{equation}  
\int_K^\infty P(k)dk={1\over N},
\end {equation}
yielding $K\approx mN^{1/(\alpha-1)}$. (For the Internet, $m=1$ and $K\approx 
N^{2/3}$.) For the sake of generality, below we consider a range of variables,
 $\alpha\geq 1$ and $1\leq m\ll K$. The key parameter, according to~
(\ref{perc}), is the ratio of second- to first-moment, $\kappa_0$, which we 
compute by approximating the distribution~(\ref {SF}) to a continuum (this 
approximation becomes exact for $1\ll m \ll K$, and it preserves the essential
 features of the transition even for small $m$):
\begin{equation}
\label{SF_perc}
\kappa_0= \biggl({2-\alpha\over3-\alpha}\biggr) 
{{K^{3-\alpha}-m^{3-\alpha}}\over{K^{2-\alpha}-m^{2-\alpha}}}.
\end{equation}
When $K\gg m$, this may be approximated as:
\begin{equation}
\kappa_0\rightarrow \biggl|{2-\alpha\over 3-\alpha}\biggr|
\times\begin{cases} m,& \text{if $\alpha>3$;}\\ 
m^{\alpha-2}K^{3-\alpha}, &\text{if $2<\alpha<3$;}\\ 
K,& \text{if $1<\alpha<2$.}
\end {cases}
\end {equation}
We see that for $\alpha> 3$ the ratio $\kappa_0$ is finite and there is a 
percolation transition at 
$1-\pc=\bigl({\alpha-2\over \alpha-3} m -1\bigr)^{-1}$: for $p>\pc$ the 
spanning cluster is fragmented and the network is destroyed.  However, for 
$\alpha<3$ the ratio $\kappa_0$ diverges with $K$ and so $\pc\rightarrow 1$ 
when $K\rightarrow \infty$ (or $N\rightarrow \infty$). The percolation 
transition does not take place: a spanning cluster exists for arbitrarily large
 fractions of breakdown, $p<1$. In {\it finite} systems a transition is always
observed, though for $\alpha<3$ the transition threshold is exceedingly high. 
For the case of the Internet ($\alpha\approx 5/2$), we have
$\kappa_0\approx K^{1/2} \approx N^{1/3}$. Considering the enormous size of the
 Internet, $N>10^6$, one needs to destroy over 99\% of the nodes before the 
spanning cluster collapses.

\begin{figure}   
\narrowtext \epsfxsize=2.7in 
\hskip 0.5in
\epsfbox{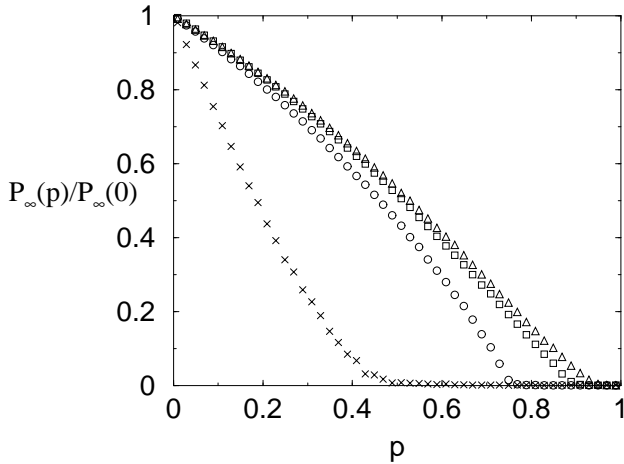} 
\vskip 0.15in
\caption{ Percolation transition for networks with power-law connectivity 
distribution. Plotted is the fraction of nodes that remain in the spanning 
cluster after breakdown of a fraction $p$ of all nodes, 
$P_\infty (p)/P_\infty (0)$, as a function of $p$, for $\alpha=3.5$(crosses) 
and $\alpha=2.5$ (other symbols), as obtained from computer simulations of up 
to $N=10^6$. In the former case, it can be seen that for $p>\pc\approx 0.5$ the
 spanning cluster disintegrates and the network becomes fragmented. However, 
for $\alpha=2.5$ (the case of the Internet), the spanning cluster persists up 
to nearly 100\% breakdown. The different curves for $K=25$ (circles), $100$ 
(squares), and $400$ (triangles) illustrate the finite size-effect: the 
transition exists only for finite networks, while the critical threshold $\pc$
 approaches 100\% as the networks grow in size.
\label{fig1}}
\end{figure}

The transition is illustrated by the computer simulation results shown in Fig.
 \ref{fig1}, where we plot the fraction of nodes which remain in the spanning 
cluster, $P_\infty (p)/P_\infty (0)$, as a function of the fraction of random 
breakdown, $p$, for networks with the distribution (\ref{SF}). For 
$\alpha=3.5$, the transition is clearly visible: beyond $\pc\approx 0.5$ the 
spanning cluster collapses and $P_\infty(p)/P_\infty (0)$ is nearly zero. On 
the other hand, the plots for $\alpha=2.5$ (the case of the Internet) show that
 although the spanning cluster is diluted as $p$ increases 
($P_\infty (p)/P_\infty (0)$ becomes smaller), it remains connected even at 
near 100\% breakdown. Data for several system sizes illustrate the finite-size
 effect: the transition occurs at higher values of $p$ the larger the simulated
 network. The Internet size is comparable to our largest simulation, making it 
remarkably resilient to random breakdown. 

We have introduced a general criterion for the collapse of randomly connected 
networks under random removal of their nodes. This criterion, when applied to 
the Internet, shows that the Internet is resilient to random breakdown of its 
nodes: a cluster of interconnected sites which spans the whole Internet becomes
 more dilute with increasing breakdowns, but it remains essentially connected 
even for nearly 100\% breakdown. The same is true for other networks whose 
connectivity distribution is approximately described by a power-law, as in 
Eq.~(\ref{SF}), as long as $\alpha<3$.

\acknowledgments  
After completing this manuscript we learned that Eqs.~(\ref{criterion}) and 
(\ref{kappa}) have been derived earlier using a different approach by Molloy 
and Reed \cite{molloy}. We thank Dr. Mark E. J. Newman for bringing this 
reference to our attention. We thank the National Science Foundation for 
support, under grant PHY-9820569(D.b.-A.).

\end{multicols}
\end{document}